**First Steps, Lasting Impact: Platform-Aware Forensics for the Next Generation of Analysts**

**Vinayak Jain [a], Sneha Sudhakaran [a], Saranyan Senthivel [b]**
[a]Florida Institute of Technology, Melbourne, USA
[b]DXC Technology
vjain2023@my.fit.edu
ssudhakaran@fit.edu
ssenthivel2@dxc.com

**Abstract:** The reliability of cyber forensic evidence acquisition is strongly influenced by the underlying operating systems, Windows, macOS, and Linux - due to inherent variations in file system structures, encryption protocols, and forensic tool compatibility. Disk forensics, one of the most widely used techniques in digital investigations, faces distinct obstacles on each platform. Windows, with its predominantly NTFS and FAT file systems, typically supports reliable disk imaging and analysis through established tools such as FTK Imager and Autopsy/Sleuth Kit. However, encryption features frequently pose challenges to evidence acquisition. Conversely, Linux environments, which rely on file systems like ext4 and XFS, generally offer greater transparency, yet the transient nature of log retention often complicates forensic analysis. In instances where anti-forensic strategies—such as encryption and compression— render traditional disk forensics insufficient, memory forensics becomes crucial. While memory forensic methodologies demonstrate robustness across Windows and Linux platforms forms through frameworks like Volatility, platform-specific difficulties persist. Memory analysis on Linux systems benefits from tools like LiME, snapshot utilities, and dd for memory acquisition; nevertheless, live memory acquisition on Linux can still present challenges. This research systematically assesses both disk and memory forensic acquisition techniques across samples representing Windows and Linux systems. By identifying effective combinations of forensic tools and configurations tailored to each operating system, the study aims to improve the accuracy and reliability of evidence collection. It further evaluates current forensic tools and highlights a persistent gap: consistently assuring forensic input reliability and footprint integrity. By integrating static analysis with temporal performance evaluations, this study emphasises the importance of developing standardized forensic procedures to ensure the integrity and admissibility of evidence across diverse platforms. Establishing such standardised methodologies is imperative to maintaining the credibility and reliability of cyber forensic investigations in heterogeneous operating system environments.

**Keywords:** digital forensics; evidence acquisition; memory forensics; disk forensics; anti-forensic techniques; forensic tool reliability

**1. Introduction**

The reliability of cyber forensic evidence acquisition is strongly influenced by the operating system under investigation, specifically Windows, macOS, or Linux, due to fundamental differences in file system structures, encryption mechanisms, and the compatibility of forensic tools. Among the most widely adopted approaches, disk forensics provides a foundation for digital investigations; however, it encounters distinct challenges across various platforms. Windows systems, which predominantly employ NTFS and FAT file systems, are well supported by established tools such as FTK Imager (FTK Imager, 2000) and Autopsy/Sleuthkit(Autopsy,2001; Sleuthkit,2001) . Still, built-in encryption features often complicate the acquisition of reliable information. Linux environments, in contrast, commonly use ext4 and XFS file systems that offer greater transparency. However, the ephemeral nature of system logs and varying retention policies creates obstacles to comprehensive evidence collection. When traditional disk-based approaches are hindered by encryption, compression, or other anti-forensic techniques, memory forensics becomes indispensable. Memory acquisition and analysis frameworks, such as Volatility, offer cross-platform capabilities; however, platform-specific hurdles persist. On Linux, for example, tools such as LiME (LIME, 2012)] and raw memory acquisition with dd support forensic collection, though live memory acquisition continues to pose technical difficulties(Forensics College, 2023).

This research systematically examines both disk and memory forensic acquisition techniques across representative Windows and Linux systems. By identifying the most effective combinations of tools and configurations tailored to each platform, the study seeks to enhance the accuracy and reliability of digital evidence collection. Beyond evaluating practical performance of forensic tools, it highlights a persistent research gap: ensuring the consistency, reliability, and integrity of forensic inputs across heterogeneous environments. To address this challenge, the study integrates static analysis with temporal performance assessments, underscoring the need for standardized forensic procedures. Such standardization is critical to safeguarding the credibility and admissibility of cyber forensic evidence in contemporary investigations.

Detecting anti-forensic techniques ensures that attempts at obfuscation or deletion do not obscure critical evidence. Building on the preceding analysis, we aim to identify reliable and effective forensic methods by conducting an empirical evaluation across a range of simulated case studies. Our contribution in this paper marks two primary research objectives:

**1. Research Objective 1**: To systematically evaluate the effectiveness of different forensic techniques across simulated case studies, identifying those that yield the most reliable and trustworthy results.

**2. Research Objective 2**:To develop a set of best practices for assessing the credibility and accuracy of forensic methods through empirical analysis, ensuring consistency in real world applications.

In this paper, Section 2 explains the related work, Section 3 describes the design methodology, Section 4 presents results.

**2. Literature Survey**
**2.1 Cross-Platform Forensic Analysis for Cybercrime Investigations**
Digital, computer, and cyber forensics have emerged as critical disciplines in modern criminal investigations, driven by the widespread proliferation of computing devices. This growth has introduced both opportunities and challenges for investigators. As cybercrime becomes increasingly sophisticated, forensic methodologies must evolve to address the diverse array of operating systems encountered in investigations. This literature survey focuses on empirical approaches to forensic analysis across desktop operating systems – Windows, Linux, and macOS, for future work reference, with a specific emphasis on the challenges posed by malware infections. The importance of cross-platform forensic proficiency has increased significantly as
Cyber criminals increasingly employ multiple operating systems in their operations. Recent findings suggest that approximately 43% of sophisticated cyberattacks involve multiple operating systems, either as targets or as platforms for attack (EBSCO, 2023). This multi-platform strategy presents distinct challenges for forensic investigators, who must possess expertise across various operating environments, each with unique file systems, memory management practices, and evidence artifact locations. This survey synthesizes current research, methodological frameworks, and empirical findings to provide a comprehensive understanding of reliable forensic analysis techniques applicable across diverse operating systems. Particular emphasis is placed on memory dumps and disk images as primary evidence sources, with close examination of how the presence of malware impacts the acquisition and analysis of these forensic artifacts. Commercial forensic tools are widely used for evidence acquisition and analysis, with varying cross-platform capabilities. For instance, EnCase and X-Ways Forensics are regarded for windows( registry artifacts, NTFS), while BlackLight targets APFS on macOS and Magnet AXIOM provides cross-platform timeline analysis (ForensicColleges, 2023;Magnet,2024). One common drawback of commercial forensic tools is their delay in supporting new operating system versions. Studies have shown that it often takes several months for these tools to fully adapt to major updates, with macOS updates typically

---

experiencing the longest delays compared to Windows and Linux distributions. This delay can impede timely investigations, particularly when dealing with cutting-edge technologies or recently released operating systems (Forensic Focus, 2023).

### 2.1.1 Open-Source Forensic Tools

Open-source forensic tools are becoming increasingly popular as cost-effective and flexible alternatives to commercial solutions. Tools such as The Sleuth Kit offer strong support for Linux file systems like EXT4, though they often lack comprehensive compatibility with macOS's APFS(Sleuthkit, 2001). In cloud environments, Google Cloud Forensics Utils integrates open-source frameworks such as Plaso and Log2Timeline to streamline log analysis in hybrid environments(CloudForensicutil, 2023). For memory forensics, Volatility has established itself as a leading framework, providing robust capabilities for examining memory artifacts on both Windows and Linux platforms (Ligh,2014;Volatility,2014). Open-source solutions frequently demonstrate faster adaptation to emerging technologies than their commercial counterparts, driven by active community contributions.

This adaptability makes them particularly valuable for investigating unusual configurations, addressing novel file systems, and responding to newly developed platforms (BlueVoyant, 2025 ; ForensicFocus, 2023). The above literature has helped us to inform tool selection for acquisition and analysis across platforms. Digital forensic investigations are based on established frameworks to ensure systematic and methodical handling of evidence. Among the most prominent is the DFRWS Investigative Model, which comprises six well-defined phases: identification, preservation, collection, examination, analysis, and presentation (Palmer, 2001). This framework emphasizes scientific rigor and legal admissibility, serving as a cornerstone for modern investigative methodologies (Infosec, 2016). Furthermore, the ACPO Good Practice Guide builds on these phases, placing particular emphasis on the chain-of-custody protocols and the validation of forensic tools (Colvee,2023). There are many other tools that are specific to mobile device forensics like AmpleDroid, DroidScraper(Ali-Gombe et al, 2017; Ali-Gomber et al,2023; Sudhakaran,2022; Sudhakaran et al,2022; Sudhakaran et al, 2020). Though there are multiple open source tools developed in forensics the creation of such tools are also to be made aware to public as we see that there are surveys conducted which reveal that usage of such tools for people from academia is not easy and is not much accessible as shown in the work presented in(Sudhakaran et al, 2024).

### 2.2 Acquisition of Evidential Data in a Trusted Environment

Once we identify that the forensically acquired dump has a role in anti-forensic data,a forensic investigator must acquire runtime or volatile data. This enables the effective acquisition and analysis of volatile data for efficient evidence recovery. The acquisition of volatile data is one of the most critical and technically challenging steps in digital forensics. Preserving ephemeral evidence before it is lost is paramount, and memory forensics plays a central role in this effort. By analyzing system RAM, investigators can recover valuable information that is not present in persistent media. Tools such as Belkasoft Live RAM Capturer are widely used in Windows systems due to their ability to operate at the kernel level and bypass anti-debugging or anti-dumping measures ensuring reliable memory extraction (ForensicsFocus,2023; Belkasoft,2023).

For Linux, LiME (Linux Memory Extractor) provides full memory captures while limiting interactions between user and kernel space processes, thus maintaining forensic soundness. However, LiME can face challenges when compiling kernel modules in various Linux distributions [48, 49]. On macOS, tools such as OSXPMem and MacQuisition are commonly used for capturing volatile memory. Although effective, these tools
can encounter difficulties with Apple's advanced memory protection features, sometimes resulting in system crashes that risk data loss if not properly managed (Naval,2017;Ponder,2017). Delays in acquisition increase the chance of losing critical artifacts, as even minimal tool activity can modify memory contents. Despite these complications, volatile memory analysis remains indispensable, often uncovering encryption keys, login credentials, and other vital evidence Forensicswiki, 2023;Belkasoft,2023; Ponder,2017).

---

Hash verification is a cornerstone of verifying evidence integrity. Although MD5 and SHA-1 are still widely used, studies have highlighted their susceptibility to collision attacks, prompting a shift towards stronger hashing algorithms, such as SHA-256 (TechFusion,2024). Adopting these more robust algorithms is recommended for verifying all digital evidence, regardless of the operating system's source. Specialized forensic distributions, including SANS SIFT, Kali Linux, and CAINE, have gained prominence for creating standardized analysis environments. Each distribution offers unique strengths. SIFT is known for its broad compatibility with Windows, Linux, and macOS, while CAINE excels at integrating legal documentation workflows, but provides limited support for macOS artifacts. Cloud-based platforms are emerging as essential tools in digital investigations. Solutions like Cado Security automate evidence collection in cloud environments such as AWS and Azure, significantly reducing analysis time compared to traditional methods (APUS,2024; Sentinel,2024). Timeline reconstruction tools, such as Magnet AXIOM, correlate events across platforms to establish comprehensive attack chronologies. Challenges such as clock synchronization discrepancies between systems (e.g., Windows vs. Linux) complicate cross-platform timelines but can be mitigated through rigorous NTP implementation. Visualization techniques such as temporal heat maps improve the recognition of attack patterns, reducing analysis time, and enhancing investigative efficiency (SalvationData,2023). Correlating digital artifacts across platforms is essential for modern investigations. Tools such as Cado Security map cloud activity to on-premises evidence, linking AWS/Azure logs with local user actions. Graph-based analysis models files, processes, and users as nodes with interactions as edges, revealing hidden connections in multi-platform attacks. Behavior-based correlation focuses on attacker objectives rather than platform-specific methods, improving the detection of advanced threats tailored to specific environments (CadoSecurity,2023).

## 3. Design Methodology

This part presents the empirical framework adopted to evaluate the reliability of forensic methods across Windows and Linux platforms. The methodology as shown in Fig 1 is structured into three modules: Case Study Creation, Reliable Evidence Acquisition, and Reliable Analysis for Investigation.

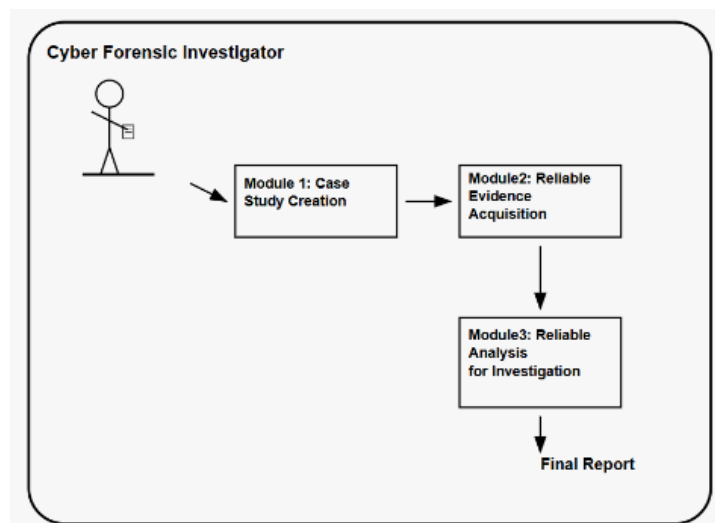

**Figure 1: Environment Setup**

### 3.1 Module 1 – Case Study Creation

All experiments were performed in standardized virtual environments (VMware and VirtualBox) on an Intel i5-7200U host system to minimize external artifacts. Four operating systems—Windows 11, Windows 10, Ubuntu 24, and Ubuntu 20—were configured with 4GB and 8GB RAM and 80GB storage and is depicted in Fig 2. Simulated user activity (browsing, file handling, and media use) created realistic conditions. Baseline clean-state memory and disk images were collected for comparison.

| Operating System | Version | RAM Configuration | Storage Capacity |
|---|---|---|---|
| Windows | Windows 11 | 4GB / 8GB | 80GB |
| Windows | Windows 10 | 4GB / 8GB | 80GB |
| Linux | Ubuntu 24 | 4GB / 8GB | 80GB |
| Linux | Ubuntu 20 | 4GB / 8GB | 80GB |

**Figure 2: Operating Systems and Their Configurations**

A total of 32 case studies per platform were designed to evaluate forensic reliability under increasing infection complexity. Malware included Trojans (CrossRAT, veryfun.exe, icon_dance.exe; dakkatoni, gosh, brootkit) and ransomware (cerber, cryptolocker, WannaCry; ransomexx, monti, gonnacry) from open Internet randomly selected. Infection states ranged from clean to single, double, and triple infections, allowing for the analysis of evidence persistence and malware evolution.

**3.2 Module 2 – Reliable Evidence Acquisition**

**Stage C0:** Each malware type was installed on a clean VM and executed for at least 10minutes (longer for complex samples) before acquisition. **Stage C1:** One malware instance (same category) was executed, then acquired. **Stage C2:** Two instances (same category) were executed, then acquired. **Stage C3:** Three instances were executed in isolated environments, then acquired to prevent spread while maintaining realism. During these stages, systems remained operational with regular user-like activities, including web browsing, document file access, and application launches and closures. This approach allowed for the assessment of the
persistence of evidence and the evolution of malware behavior over time in different operating systems. Memory dumps were collected with FTK Imager (Windows) and LiME (Linux). Disk images were acquired with FTK Imager (Windows) and dd (Linux). Each acquisition was repeated under both RAM configurations, and cryptographic hashes were used to ensure the integrity of the evidence.
Sections 3.1 and 3.2 satisfy Research Objective 1.

**3.Module 3 – Reliable Analysis for Investigation**

This experimental study provides a systematic set of best practices for evaluating forensic capabilities in diverse computing environments. By implementing various test scenarios across different operating systems with varying hardware configurations, the research enables empirical assessment of the effectiveness of forensic tools in real-world scenarios. We created analysis scenarios for trojans and ransomware and evaluated their effectiveness under time- and resource-dependent conditions across various tools. We conducted experiments with multiple time and resource-dependent analyses across different tools and ensured the scenarios were achieved. The Section 3.3 satisfies Research Objective 2

*3.3.1. Malware-specific analysis challenge*
Trojans may hide processes, inject into legitimate processes, and establish persistent connections detectable in memory forensics but harder to observe in disk forensics. Ransomware will likely create readily identifiable artifacts in both disk and memory forensics due to extensive file system modifications and encryption operations.

*3.3.2. Operating system-specific challenges may include*

Windows systems exhibited complications with registry analysis and memory address translation. Linux systems presented challenges with different kernel versions affecting memory structure mapping.

**4. Evaluation and Effectiveness Assessment**

The empirical analysis highlights the impact of trojans and ransomware on Windows and Linux systems. We conducted memory analysis with Volatility on dumps acquired from both platforms.

**4.1 Windows Case Study Findings: Impact of Trojans and Ransomware on Forensic Analysis**

*4.1.1 Memory Forensic Analysis on Trojan and Ransomware Case Investigations*

| Volatility plugin | C0 Clean state | | C1 ransomware infection | | C2 ransomware infection | | C3 ransomware infection | |
|---|---|---|---|---|---|---|---|---|
| | 8gb | 4gb | 8gb | 4gb | 8gb | 4gb | 8gb | 4gb |
| Windows.pslist | 131 | 133 | 132 | 141 | 125 | 136 | 120 | 122 |
| Windows.psscan | 132 | 139 | 131 | 143 | 135 | 141 | 123 | 144 |
| Windows.pstree | 130 | 133 | 132 | 140 | 133 | 138 | 120 | 135 |
| Windows.malfind | 22 | 24 | 14 | 11 | 19 | 11 | 5 | 7 |

**Figure 3: Windows 10 ransomware malware**

Based on these findings in Fig 3 and 4, we recommend prioritizing 8 GB Windows 10/11 environments for memory forensics in Trojan-infected systems to maximize evidence retention and analysis efficiency, while acknowledging the potential but limited utility of 4GB systems where hardware constraints apply. Greater memory capacity improves artifact retention and accommodates system-level constraints such as page swapping.

*4.1.2 Disk Forensic analysis on Trojan and ransomware Case Investigation*

| Analysis results | C0 clean state | C1 Trojan infection | C2 Trojan infection | C3 Trojan infection | C1 Ransomware | C2 Ransomware | C3 Ransomware |
|---|---|---|---|---|---|---|---|
| Encryption Suspected | 4 | 3 | 2 | 2 | 11 | 9 | 8 |
| Extension Mismatch Detected | 254 | 261 | 260 | 265 | 255 | 383 | 215 |
| Suspicious items | 258 | 264 | 263 | 268 | 267 | 393 | 223 |
| Bad items | 0 | 0 | 0 | 0 | 0 | 0 | 0 |
| Interested Item (zip bomb) | 0 | 0 | 1 | 1 | 1 | 1 | 0 |
| Recycle bin | 19 | 19 | 19 | 19 | 19 | 19 | 19 |

**Figure 4: Analysis Results of Various Infections of Windows 10 (8 GB)**

Case studies involving Trojan and ransomware infections revealed significant variations across Windows versions. We conducted the analysis using Autopsy 4.21.0, focusing on suspected encryption activities and extension mismatches, as well as identifying suspicious items, bad items, potential zip bombs, and artifacts within the Recycle Bin. Based on empirical results, we conclude that Windows 11 demonstrates a stronger capability to resist and detect malware compared to Windows 10 within the context of our disk forensics analysis. We observed a slight increase in detected artifacts from the C1 to C3 malware infection states, except for the C3 ransomware infection stage.

**4.2 Linux Case Study Findings: Impact of Trojans and Ransomware on Forensic Analysis**

*4.2.1 Memory Forensic Findings Across Linux Systems*

Examining memory dumps from various Linux environments infected with trojan malware revealed notable differences in forensic visibility as shown in Figures 5 and 6. We conducted our analysis using the Volatility 3 framework with Linux-specific plugins, including linux.pslist, linux.pstree, and linux.pidhashtable, to identify distinct malware behaviors. Additionally, we utilized the linux.elf plugin to detect suspicious or malicious ELF files during memory forensics.

| Volatility plugin | C 0 Clean state | | C1 trojan infection | | C2 trojan infection | | C3 trojan infection | |
|---|---|---|---|---|---|---|---|---|
| | 8gb | 4gb | 8gb | 4gb | 8gb | 4gb | 8gb | 4gb |
| linux.pslist | 148 | 153 | 151 | 137 | 136 | 129 | 145 | 137 |
| linux.pidhashtable | 151 | 155 | 140 | 143 | 143 | 144 | 152 | 136 |
| linux.pstree | 147 | 153 | 149 | 91 | 136 | 121 | 150 | 139 |

**Figure 5 : Ubuntu 24 trojan malware**

Figure 5 above presents the number of processes identified by each plugin during the analysis of Ubuntu 24.04 and Ubuntu 20.04 systems with 8 GB and 4 GB RAM configurations infected with trojan malware. Our analysis revealed that trojan malware infections exhibited similar behavioural patterns in both Linux operating systems examined. However, we observed a notable difference in the time required to collect memory dumps between the systems. Specifically, Ubuntu 24.04 required, on average, 5–8% longer to complete memory dump acquisition compared to Ubuntu 20.04.

| Volatility plugin | c 0 Clean state | | C1 ransomware infection | | C2 ransomware infection | | C3 transomware infection | |
|---|---|---|---|---|---|---|---|---|
| | 8gb | 4gb | 8gb | 4gb | 8gb | 4gb | 8gb | 4gb |
| linux.pslist | 144 | 134 | 154 | 112 | x | x | x | x |
| linux.pidhashtable | 150 | 138 | 151 | 134 | x | x | x | x |
| linux.pstree | 148 | 138 | 154 | 115 | x | x | x | x |

**Figure 6 : Ubuntu 24 Ransomware malware**

Figures 5 and 6 clearly shows the ransomware infections created substantially different forensic patterns compared to trojans. We have used the same procedure to take the memory dumps as in the trojan infection case and the same volatility plugins for memory forensic analysis. Our observations indicate that ransomware variants such as monti and gonnacry exhibit a significantly higher potential to harm systems compared to trojan malware. Ransomware infections result in a greater degree of system corruption, actively modifying system files in ways that can interfere with the memory acquisition process, whereas trojan infections generally do not cause such extensive file system alterations.

*4.2.2 Disk Forensic Findings Across Linux Systems*

Based on our analysis across both system configurations (8GB and 4GB RAM) running the same Linux operating system, we found that variations in RAM capacity did not produce a substantial difference in the results of disk forensics. Compared to trojan malware, ransomware infections in Linux systems exhibited an approximate 8–15% increase in suspicious activity. trojan infections, by contrast, showed a 3–5% increase in activity as the number of trojan samples introduced into the system increased. Analysis of the Recycle Bin revealed that, even after ransomware encryption, disk forensics remained effective in recovering deleted data artifacts from the Recycle Bin.

### 4.3 Errors and Complications in Forensic Analysis

During our comprehensive analysis across operating systems, we encountered various errors and complications that impacted forensic processes and outcomes. These challenges arose across both Windows and Linux environments and were influenced by factors related to the host machines used for case studies as well as the virtual machine platforms employed, including VMware Workstation and Oracle VirtualBox.

*4.3.1 Operating System Complications*

Many types of malware fail to execute due to compatibility issues between the operating system and the malware's executable format. For example, Linux systems support ELF (Executable and Linkable Format) files, while Windows systems support .exe files. This disparity makes it challenging to generalize how the same malware type affects different operating systems.

On Windows 11, after acquiring forensic data on an 8 GB RAM configuration for trojan and ransomware samples, the virtual machine displayed errors such as triggering the automatic repair process. Even after several hours, the Windows 11 VM host failed to restart. Attempts to resolve the issue using a new ISO image
and switching to Oracle VirtualBox were unsuccessful, with the Windows 11 VM failing to start in both VMware Workstation and VirtualBox. Consequently, we could not acquire forensic data for the 8 GB configuration on Windows 11, an issue that did not occur on Windows 10 VMs. The host system occasionally experienced sudden crashes during forensic acquisition and analysis, manifesting as a blank white screen or system lockup/freezes.

Extended periods without restarting the host machine led to delays when opening virtual machines on both VMware Workstation and VirtualBox. Ubuntu 24.04, in particular, exhibited signs of system overload, failing to open immediately as Ubuntu 20.04 did. In these cases, restarting the
entire host machine was necessary to resume work on Ubuntu 24.04, whereas for Ubuntu 20.04, closing and reopening the VM itself was sufficient.

*4.3.2 Memory Forensic Complications*

Volatility is available as Versions 2 and 3. Because Volatility 2 does not support Windows 11, we used Volatility 3 for all Windows 10/11 memory analyses.

When analyzing memory dumps from Windows 11, Volatility may display symbol table and profile errors. If using an older version, such as Volatility 2.7.0, upgrading to the latest version is recommended to resolve these errors. It is essential to ensure that all required Python libraries are installed in the environment otherwise, ModuleNotFoundError may occur. Additionally, specifying the correct path to the memory dump is critical, as incorrect paths will result in FileNotFoundError.

Malware infections, particularly ransomware, can corrupt the system's OS symbol table. In our Linux OS case studies, we observed that at the C2 and C3 infection stages, forensic analysis often failed due to malware-induced encryption and kernel corruption, resulting in corrupted memory dumps. Consequently, Volatility was unable to process these memory dumps for analysis.

*4.3.3 Disk Forensic Complications*

In disk forensics, autopsy requires an average of 2–3 hours to process a given disk dump. During this process, it is recommended that the host machine remain idle, as concurrent activities may cause the application to freeze or enter a processing loop, necessitating a complete restart of the analysis to obtain accurate forensic data.For Windows systems, the use of FTK Imager during disk dump acquisition is recommended. This tool can export a structured list of files and directories from an acquired image to Excel-compatible formats (CSV or .xlsx), facilitating forensic reporting, timeline reconstruction, and the identification of suspicious
files, such as those with unusual extensions, large sizes, or hidden attributes.
For Linux systems, selecting the correct file-system path (e.g., /dev/sda) is essential during acquisition. Failure to specify the correct disk path will prevent Autopsy from accurately reading and processing the disk dump. Severe disk corruption or ransomware overwrites can hinder Autopsy's ability to carve and recover files from disk images successfully. Malware often hides in directories such as /tmp/.xyz or /var/.abc. These hidden files may be overlooked unless explicitly searched during analysis. Files with extensions such as .jpg or .txt may, in reality, be executable or encrypted files. Malware infections may also generate duplicate files containing random data, which share identical hash values, thereby complicating file integrity analysis.

---

*Paper is accepted at 21st International Conference on Cyber Warfare and Security (ICCWS 2026)
*Corresponding author: Sneha Sudhakaran, ssudhakaran@fit.edu

### 4.4 Limitations and Future Research Directions
While this research provides comprehensive information on the behavior of malware in Windows and Linux operating systems, several limitations must be acknowledged. The use of virtual machines, while essential for consistency and isolation, may not fully replicate physical hardware behavior(e.g., memory management, storage performance). Future studies should consider testing on physical systems to validate these findings. Furthermore, while the selected malware samples represent common cross-platform threats, they do not capture the full diversity of malware types, particularly those designed for specific operating systems. Future research should incorporate a broader range of targeted malware samples to enhance generalizability. Reliance on Volatility 3 and Autopsy 4.21.0 introduces potential tool-specific biases. Comparing results across multiple forensic tools would help distinguish genuine platform behaviors from tool-induced artifacts, thus enhancing the robustness and reliability of forensic analysis outcomes.

### 5. Ethics Declaration:
This study was conducted in accordance with established ethical and professional standards. No human participants, personal data, or sensitive information were involved; all datasets and experiments were performed on controlled test environments or publicly available resources. The work is intended solely for research and educational purposes in cybersecurity and does not promote or support malicious use.

### 6. AI declaration:
Artificial intelligence tools (OpenAI ChatGPT, GPT-5) were employed solely to improve grammar, readability, and formatting of the manuscript text. The authors take full responsibility for the scientific content, data interpretation, and conclusions. The authors also used this tool response as this work mainly focus on the usage of Gen AI tool like Chat Gpt with Human forensic investigators for better efficacy over cyber forensic investigation. All outputs from AI tools were carefully reviewed and edited by the authors to ensure accuracy and originality.

### 7. Conclusion
This research presents a systematic empirical analysis of reliability challenges in cyber forensic investigations across Windows and Linux environments, using a controlled experimental framework with virtual machines to ensure hardware consistency. Four operating systems were evaluated under varied malware infection scenarios, system configurations, and acquisition timelines, revealing that operating system architecture, memory resources, malware type, and the timing of forensic acquisition significantly affect evidence visibility and degradation. By incorporating both memory and disk forensics with standardized procedures and platform-specific tools, the study provided consistent and credible cross-platform evaluations. Key contributions include a cross-platform forensic comparison under uniform malware conditions, quantification supporting a practical minimum of 8 GB RAM for reliable analysis; identification of distinct forensic patterns in trojan and ransomware infections, empirically grounded timelines for evidence degradation, and documentation of tool-specific error behaviors across platforms to guide practitioners. The results further suggest that encrypted content demonstrates resilience against ransomware attacks, highlighting encryption as a defensive strategy. Nevertheless, limitations arise from using virtual machines, which may not fully replicate physical hardware behaviors, the restricted set of malware samples, and potential tool-specific biases from Volatility 3 and Autopsy 4.21.0. Future work will expand this analysis to macOS environments and incorporate a broader range of targeted malware to strengthen generalizability.